\documentclass[]{aastex63}

\received{14 January 2025}
\accepted{20 January 2025}

\shorttitle{The Molecular Envelope of NGC~6720}
\shortauthors{Kastner et al.}

\begin{document}

\title{The Structure of the Molecular Envelope of the Ring Nebula (NGC~6720)}

\correspondingauthor{Joel Kastner}
\email{jhk@cis.rit.edu}

\author[0000-0002-3138-8250]{Joel H. Kastner}
\affiliation{Center for Imaging Science, 
  Rochester Institute of Technology, Rochester NY 14623, USA; jhk@cis.rit.edu}
\affiliation{School of Physics and Astronomy, 
  Rochester Institute of Technology} 
\affiliation{Laboratory for Multiwavelength Astrophysics, 
  Rochester Institute of Technology}

\author[0000-0003-1526-7587]{David J. Wilner}
\affiliation{Center for Astrophysics, Harvard \& Smithsonian, 60 Garden Street, Cambridge, MA 02138-1516, USA}

\author{Diana Ryder}
\affiliation{School of Physics and Astronomy, 
  Rochester Institute of Technology} 

\author{Paula Moraga Baez} 
\affiliation{School of Physics and Astronomy, 
  Rochester Institute of Technology} 

\author{Orsola De Marco}
\affiliation{School of Mathematical and Physical Sciences, Macquarie University, Sydney, New South Wales, Australia}
\affiliation{Astronomy, Astrophysics and Astrophotonics Research Centre, Macquarie University, Sydney, New South Wales, Australia}

\author{Raghvendra Sahai}
\affiliation{Jet Propulsion Laboratory, California Institute of Technology, 4800 Oak Grove Drive, Pasadena, CA 91109, USA}

\author{Al Wootten}
\affiliation{National Radio Astronomy Observatory, Charlottesville, VA 22903, USA}

\author{Albert Zijlstra}
\affiliation{Jodrell Bank Center for Astrophysics, 
The University of Manchester, 
Oxford Road, Manchester M13 9PL, UK}

\begin{abstract}
We present the first interferometric imaging of molecular line emission from the Ring
Nebula, NGC~6720, in the form of Submillimeter Array (SMA) observations of CO
$J=2\rightarrow 1$ emission. The SMA $^{12}$CO(2--1) mapping data, with
$\sim$3$''$ spatial resolution and 2 km s$^{-1}$ velocity resolution,
provide an unprecedentedly detailed, 3D view of the Ring's clumpy molecular envelope.
The morphology of the velocity-integrated SMA $^{12}$CO(2--1) image
closely resembles those of near-IR H$_2$ and PAH emission
in JWST/NIRCam imaging of NGC~6720, with the molecular gas 
forming a geometrically thin layer surrounding the ionized gas
imaged by HST and JWST. A simple, geometrical model of the $^{12}$CO(2--1) data
shows that the intrinsic structure of NGC~6720's molecular envelope closely resembles a
truncated, triaxial ellipsoid that is viewed close to pole-on, and that the
dynamical age of the molecular envelope is $\sim$6000 yr. The SMA $^{12}$CO(2--1) data furthermore 
reveal that filamentary features seen projected in the Ring's interior in JWST imaging are in fact 
fast-moving polar knots or bullets with radial velocities of $\pm$45--50 km s$^{-1}$ relative to systemic, and 
that the hot progenitor star remnant is
positioned at the precise geometric center of the clumpy, ellipsoidal molecular shell.
We assert that the Ring's molecular envelope was formed via a
relatively sudden, AGB-terminating mass ejection event $\sim$6000 yr ago, and that this ellipsoidal envelope
was then punctured by fast, collimated polar
outflows resulting from interactions between the progenitor
and one or more companion stars. Such an evolutionary scenario
may describe most molecule-rich, ``Ring-like'' planetary nebulae.
\end{abstract}

\section{Introduction}

The Ring Nebula (NGC~6720) is the nearby \citep[$D = 782^{+29}_{-32}$ pc;][]{BailerJones2021} archetype of the class of molecule-rich,
``Ring-like'' planetary nebulae (PNe). 
The intrinsic (three-dimensional) structure of the Ring has been debated for over a century \citep[e.g.,][]{Keeler1899}. Some investigations of the Ring support a geometry consisting of an ellipsoidal main ring with surrounding quasi-spherical halo \citep[e.g.,][]{ODell2007,ODell2013}, while others provide evidence for a bipolar structure, wherein the main ring is the nebula's dense waist and the halos are extended bipolar lobes viewed in projection \citep[e.g.,][]{Kastner1994,Bryce1994}. Given that molecule-rich PNe like NGC 6720 are the products of axisymmetric mass loss from relatively massive ($\stackrel{>}{\sim}$2 $M_\odot$) asymptotic giant branch (AGB)  progenitors \citep[][]{Kastner:1996ab}, this debate concerning the Ring Nebula's three-dimensional geometrical structure has profound implications for the mechanisms and timescales governing mass loss from AGB and post-AGB stars. 

Projected against the sky, the Ring displays a clear nested structure. 
Its inner, ionized cavity appears nearly
circular in shape, with a radius of $\sim$25$''$ ($\sim$0.1 pc),
and is bright in forbidden lines of high-ionization species (e.g.,
[O~{\sc iii}]). This photoionized cavity is surrounded by a denser,
elliptical ($\sim$70$'' \times 55''$) ring structure seen in lower-excitation atomic species
(e.g., [N~{\sc ii}]). The main elliptical ring, in turn, is encircled by two main, very extended (150$''$ and 230$''$ diameter) 
halo structures, the larger of the two being nearly perfectly circular. The central ring and external
halo system have bright near-IR H$_2$ emission counterparts \citep{Kastner1994}. 

Recently published JWST imaging of PN NGC~6720 provides an extinction-penetrating infrared view of the nebula,
elucidating the detailed projected structures of its ionized and H$_2$ emission regions \citep{Wesson2024}. 
The JWST images resolve
the nebula's main molecular ring into $\sim$20000 individual
H$_2$-emitting knots and reveal concentric arc
structures in the nebula's immense ($\sim$1 pc radius), surrounding
H$_2$ halo. \citet{Wesson2024} argue that the system of arcs within the 
halo revealed by the JWST H$_2$ images was most likely
sculpted by an unseen companion to the hot, luminous ($T_\star \sim130$ kK,
$L_\star \sim 310$ $L_\odot$) present-day central star.   An interacting binary
scenario is further supported by the significant molecular gas
component associated with the nebula's main, bright elliptical ring;
such molecule-rich toroidal structures in PNe are likely comprised of
the remnants of AGB envelope ejecta that were focused along the
orbital plane of the nebular progenitor star and a close binary
companion during a stage of rapid, possibly binary-assisted AGB mass
loss \citep[e.g.,][and references therein]{Kastner2022,Kastner2024}.  The average angular
distance between the H$_2$ arc structures in the molecular halo of NGC
6720 implies an orbital period of 280 years for such a putative binary
consisting of AGB primary and mass-loss-shaping companion
\citep{Wesson2024}. \citet{Wesson2024} also find that the central star is offset
from the geometric center of the nebula by $\sim$1500 au, and speculate that this offset may 
serve as additional evidence of stellar multiplicity. However, such a 
large displacement of the mass-losing star from the nebular center cannot be 
ascribed to the influence of the same companion that generated the arcs.

To explain various features observed in the JWST imaging -- especially the
Ring's overall elliptical morphology and the presence of a dark lane
within its central, highly ionized region --- \citet{Wesson2024}
present a schematic structural model invoking a twin-cone bipolar geometry. In this model, which is motivated by 
inferences from optical spectroscopy of the Ring Nebula's halo and core structures by \citet{Bryce1994} and \citet{Guerrero1997}, 
respectively, the axis of symmetry of the conically shaped polar lobes is viewed at an incliation $\sim$30$^\circ$ away from pole-on. However, this basic
intrinsic geometry appears at odds with that deduced from comprehensive long-slit optical
spectroscopic and HST imaging observations by \citet{ODell2007,ODell2013}. Those studies present
evidence that the Ring's intrinsic structure is that of a
triaxial, prolate ellipsoid viewed nearly pole-on, and that the Ring's
extended halo structures are AGB wind remnants.

In many respects, the JWST imaging of the Ring presented and analyzed in \citet{Wesson2024} 
closely parallels that of ERO JWST imaging of the Southern 
Ring planetary nebula (NGC 3132) analyzed in \citet{DeMarco2022}. 
In both cases, the H$_2$ emission imaged by JWST reveals knot and arc structures in exquisite detail; however, the 
JWST near-IR H$_2$ imaging yields no nebular
kinematic information and, furthermore, the near-IR H$_2$ emission only traces the hot
($\sim$1000 K), UV-illuminated and/or shock-excited molecular gas 
that constitutes the ``tip of the iceberg'' in terms of the total
molecular gas masses of the nebulae. To ascertain the
mass distribution and energetics of the molecular
envelope of NGC~6720 on size scales approaching those probed by JWST, we 
require interferometric mm-wave maps of molecular line (i.e., CO) emission. 
As we demonstrated in the  case of NGC 3132 \citep{Kastner2024}, such  interferometric 
molecular line mapping data provide the means to elucidate the 3D structure 
and shaping history of molecule-rich PNe, and thereby to strengthen the connection 
between nebular shaping processes and a potential companion(s) orbiting the central star's AGB progenitor.

However, as was also the case for NGC 3132, the only published mm-wave molecular
line mapping of NGC~6720 prior to JWST imaging consisted of single-dish CO raster maps
obtained over 30 years ago. The CO maps of NGC~6720 were obtained with the IRAM 30-meter and CSO 10-meter
telescopes, with 10$''$ sampling and beamwidths of $\sim$12$''$ and
$\sim$20$''$, respectively \citep{Bachiller1989,Stanek1995}. These single-dish IRAM and CSO mapping
observations revealed strong CO emission from the Ring's main ring, indicating its intrinsic geometry is indeed that of a ring or cylinder. 
Subsequent single-dish radio molecular line spectroscopy has demonstrated
the Ring Nebula's chemical richness \citep{Bachiller1997, Bublitz2019}, but sheds no additional light on the structure or kinematics of the PN's molecular envelope.

In this paper, we report the first interferometric molecular line observations observations of
the Ring Nebula, in the form of $\sim$3$''$ resolution Submillimeter Array (SMA)
observations of NGC~6720 in the $J=2\rightarrow 1$ rotational transitions of
$^{12}$CO and isotopologues ($^{13}$CO, C$^{18}$O) and adjacent (1.3 mm) continuum.  The SMA maps
reveal the detailed density distribution and
kinematics of cold molecular gas within NGC~6720. 
As in our recent, first-of-their-kind interferometric observations of CO in NGC 3132 with the SMA \citep{Kastner2024}, 
the new SMA CO mapping of NGC~6720 thereby provides unique 
insight into the nebula's 3D structure and shaping processes. 

\section{Observations}\label{sec:obs}

\begin{table}
\centering
\caption{\sc SMA observations of NGC~6720}
\begin{tabular}{c c c c c c c c} 
\hline
Date & Config & \# Ants & $\tau_{225~{\rm GHz}}$ & HA range & \multicolumn{3}{c}{Calibrators}  \\
~    & ~      &         & ~                      & ~        & Gain & Passband & Flux \\ 
\hline
19 Mar 2024 & SUB & 6 & 0.04 & $-3.5,+2.8$ & J1734+389,J1848+327 & 3C279 & Pallas \\
06 May 2024 & COM & 7 & 0.27 & $-5.3,-0.9$ & J1734+389,J1848+327 & 3C279 & Pallas \\
05 Jun 2024 & COM & 7 & 0.10 & $-5.3,+5.2$ & J1734+389,J1848+327 & 3C279 & Pallas \\ 
26 Jul 2024 & SUB & 7 & 0.10 & $-4.0,+4.0$ & J1734+389,J1848+327 & 3C279,BL Lac & Pallas \\
08 Sep 2024 & EXT & 7 & 0.09 & $-2.2,+5.3$ & J1734+389,J1848+327 & 3C84 & Pallas \\
\hline
\end{tabular}
\label{tab:observations}
\end{table}

\begin{table}
\centering
\caption{\sc SMA NGC~6720 molecular line and continuum image properties}
\begin{tabular}{c c c c c}
\hline 
Molecule & Rest freq. & Beam size              & Beam PA &  rms noise$^a$ \\
~        & (GHz)      & ($\arcsec\times\arcsec$) & ($\degr$) & (mJy~beam$^{-1}$) \\
\hline
$^{12}$CO $J=2\rightarrow 1$ & 230.538000  & 2.62$\times$2.36 & $84\degr$  & 48 \\
$^{13}$CO $J=2\rightarrow 1$ & 220.398684  & 3.74$\times$3.24 & $77\degr$  & 38 \\
C$^{18}$O $J=2\rightarrow 1$ & 219.560354  & 3.75$\times$3.25 & $78\degr$  & 39 \\
continuum       & 230.5       & 3.81$\times$3.30 & $78\degr$  & 0.66 \\
\hline
\end{tabular}

a) Noise (rms) as measured in 2~km~s$^{-1}$ bandwidths for spectral lines and 20~GHz bandwidth for continuum.
\label{tab:imagecubes}
\end{table}

We observed NGC~6720 with the Submillimeter Array (SMA) on Maunakea, Hawaii in its subcompact, compact, and extended 
configurations during the spring/summer of 2024. Table~\ref{tab:observations} provides a summary of these observations, 
including the observing dates, number of antennas in operation, median 225~GHz atmospheric opacities, usable hour angle ranges, 
and calibrator sources. The baseline lengths ranged from 6~m to 220~m.
The two dual-sideband receivers, each with an IF range of 4-16~GHz, 
were tuned to an LO frequency of 225.538. The resulting frequency coverage included the $^{12}$CO, $^{13}$CO, 
and C$^{18}$O $J=2\rightarrow 1$ lines that are the focus here. 
The SWARM digital backend provided 140~kHz channel spacing, corresponding to 0.18 km~s$^{-1}$ at the 230.538~GHz 
frequency of the $^{12}$CO $J=2\rightarrow 1$ line. We observed NGC~6720 over a hexagonal mosaic of 19 pointings with $30''$ spacing.
This setup covered a region of approximately $180\arcsec\times180\arcsec$ in size with uniform sensitivity, 
spanning the full extent of $^{12}$CO emission previously imaged with the IRAM 30~m telescope \citep{Bachiller1989}.
The SMA primary beam FWHM is $55''$ at the $^{12}$CO $J=2\rightarrow 1$ line frequency. 
The center position of the mosaic was 18$^h$53$^m$35\fs0967, $+33\degr01\arcmin44\farcs9$ (J2000).
The observing sequence consisted of $2\times10$~second integrations on each of the 19 mosaic pointings, 
bracketed by $6\times10$~second observations of the two calibrators 1734+389 and 1848+327. 
This rapid cycling provided consistent $u,v$ coverage for all of the mosaic pointings. 

For calibration, we followed standard procedures for SMA data using the MIR/IDL software package. First, the visibilities
were inspected manually and any channels that showed evidence for interference were flagged. Second, the bandpass response was 
determined from observations a strong source or sources. Third, the absolute flux scale was set from a short observation of 
the asteroid Pallas in each track (with $\sim10$\% estimated systematic uncertainty). Fourth, time dependent complex gains were 
derived and applied from observations of the calibrators interspersed through the observations. Finally, the calibrated 
visibilities were exported from MIR/IDL in a format for the MIRIAD software package for imaging. 
We used MIRIAD to make image cubes for each of the CO isotopologues using the mosaic option in the {\tt invert} task
and clean deconvolution with the {\tt mossdi} task. Synthesized beam sizes were $\sim2\arcsec - 4\arcsec$, depending on the 
visibility weighting scheme. For the $^{12}$CO $J=2\rightarrow 1$ line, we adopted robust=0 weighting as a compromise between higher angular resolution and 
better surface brightness sensitivity. For the rarer CO isotopologues, we adopted
natural weighting to obtain the best sensitivity. Each fiducial spectral line image cube has 64 velocity bins of 
2.0~km~s$^{-1}$ width that covers the full velocity range of the detected line emission (and beyond). 
We also made a continuum image, excluding the spectral windows with the CO isotopologue lines,  using natural weighting.
Table~\ref{tab:imagecubes} lists basic properties of the images, including the beam sizes and rms noise levels. 


\section{Results}\label{sec:results}

Channel maps obtained from the image cubes of $^{12}$CO and $^{13}$CO
  $J=2\rightarrow 1$ emission  are presented
  Fig.~\ref{fig:COchannelMaps} and Fig.~\ref{fig:13COchannelMaps},
  respectively. The velocity-integrated $^{12}$CO(2--1) line intensity, as measured over the velocity range $-$30 km~s$^{-1}$ to +20 km~s$^{-1}$ within a $\sim$$100'' \times 80''$ elliptical region centered on the main ring, is 1350$\pm$135 Jy km s$^{-1}$ (where the uncertainty is dominated by the estimated uncertainties in calibration; see \S~\ref{sec:obs}). The $^{13}$CO(2--1) emission is very weak, and is only detected at the velocities and locations of strongest $^{12}$CO(2--1) emission. The noise level in individual velocity channels in the $^{13}$CO(2--1) image cube is significant relative to (typically 5--10\% of) the $^{12}$CO(2--1) surface brightness at the same velocities and spatial locations, precluding a reliable measurement of the total $^{13}$CO(2--1) line intensity.  
Weak 230 GHz continuum emission is detected; as with the $^{13}$CO(2--1) line observations, the data are too noisy to obtain a reliable measurement of then total continuum flux within the band. We do not detect C$^{18}$O(2--1) emission.

The velocity-integrated
(moment 0) image of $^{12}$CO(2--1) emission and 230 GHz continuum image are displayed in
  Fig.~\ref{fig:compare_mom0H2contPAH} (left and right center panels, respectively), alongside the JWST/NIRCam 2.12
  $\mu$m H$_2$ and 3.35 $\mu$m (PAH) images (left center and right panels, respectively). 
``Collapsed views'' of the SMA $^{12}$CO(2--1) data cube, integrated in velocity (moment 0 image) and in the RA and declination dimensions (P-V images), are displayed in Fig.~\ref{fig:compare_cubes}. 
The spatially integrated $^{12}$CO(2--1) spectrum obtained from the SMA data cube, presented in Fig.~\ref{fig:COspectrum}, closely matches the ``average'' $^{12}$CO(2--1) line profile obtained via (single-dish) IRAM 30-meter raster mapping by \citet{Bachiller1989}.
  
In the SMA data, the detected $^{12}$CO(2--1) emission from NGC~6720 is confined to the central bright ring 
 (see, e.g., Figs.~\ref{fig:COchannelMaps},~\ref{fig:compare_mom0H2contPAH}). The elliptical CO emission ring has outer dimensions of $\sim$$75'' \times 55''$, or approximately $1.0\times10^{18}$ cm by $7\times10^{17}$ cm for an assumed distance of 782 pc. No CO is detected from the Ring's 
 extended halo structures in these data, despite the fact that the FOV of the SMA mosaic 
 includes the inner ($\sim$150$''$ diameter) optical/H$_2$ halo.
  
\subsection{CO emission morphology; spatial and velocity structures}\label{sec:COmorph}

The overall morphology of the CO emission from the Ring, as established in the early CO mapping \citep{Bachiller1989,Stanek1995}, is reproduced in our new SMA interferometric imaging. However, the $\sim$2.5--3$''$ resolution of the $^{12}$CO(2--1) SMA imaging represents an order of magnitude improvement on the spatial resolution of previous (single-dish) mm-wave CO mapping of NGC~6720. The moment 0 SMA image thus reveals, for the first time, the clumpy nature of the cold molecular gas within the nebula. In the integrated intensity image (Fig.~\ref{fig:compare_mom0H2contPAH}), the brightest clumps appear as distinct condensations along the major axis of the elliptical ring, with filamentary emission found along the ring's minor axis. The 230 GHz continuum image, though noisy, closely traces the $^{12}$CO(2--1) and PAH emission (Fig.~\ref{fig:compare_mom0H2contPAH}), suggesting the mm-wave continuum is likely dominated by thermal emission from cool dust embedded in the molecular gas. 

\begin{figure}[ht]
\begin{center}
\includegraphics[width=6.5in]{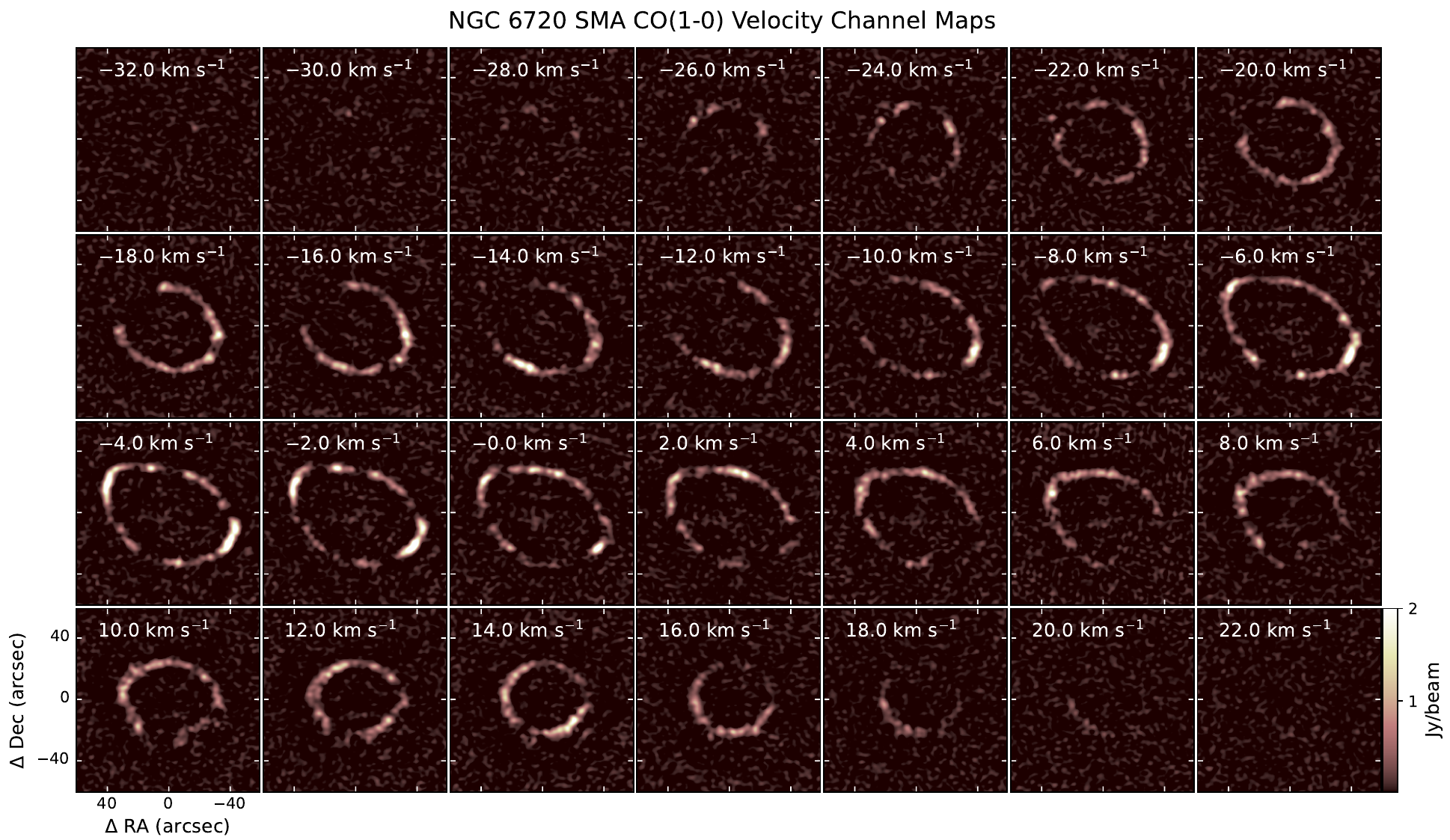}
\end{center}
\caption{SMA channel maps of $^{12}$CO(2--1) emission from NGC~6720. The synthesized beam size is 2.62$''\times$2.36$''$.}
\label{fig:COchannelMaps}
\end{figure}

\begin{figure}[ht]
\begin{center}
\includegraphics[width=6.5in]{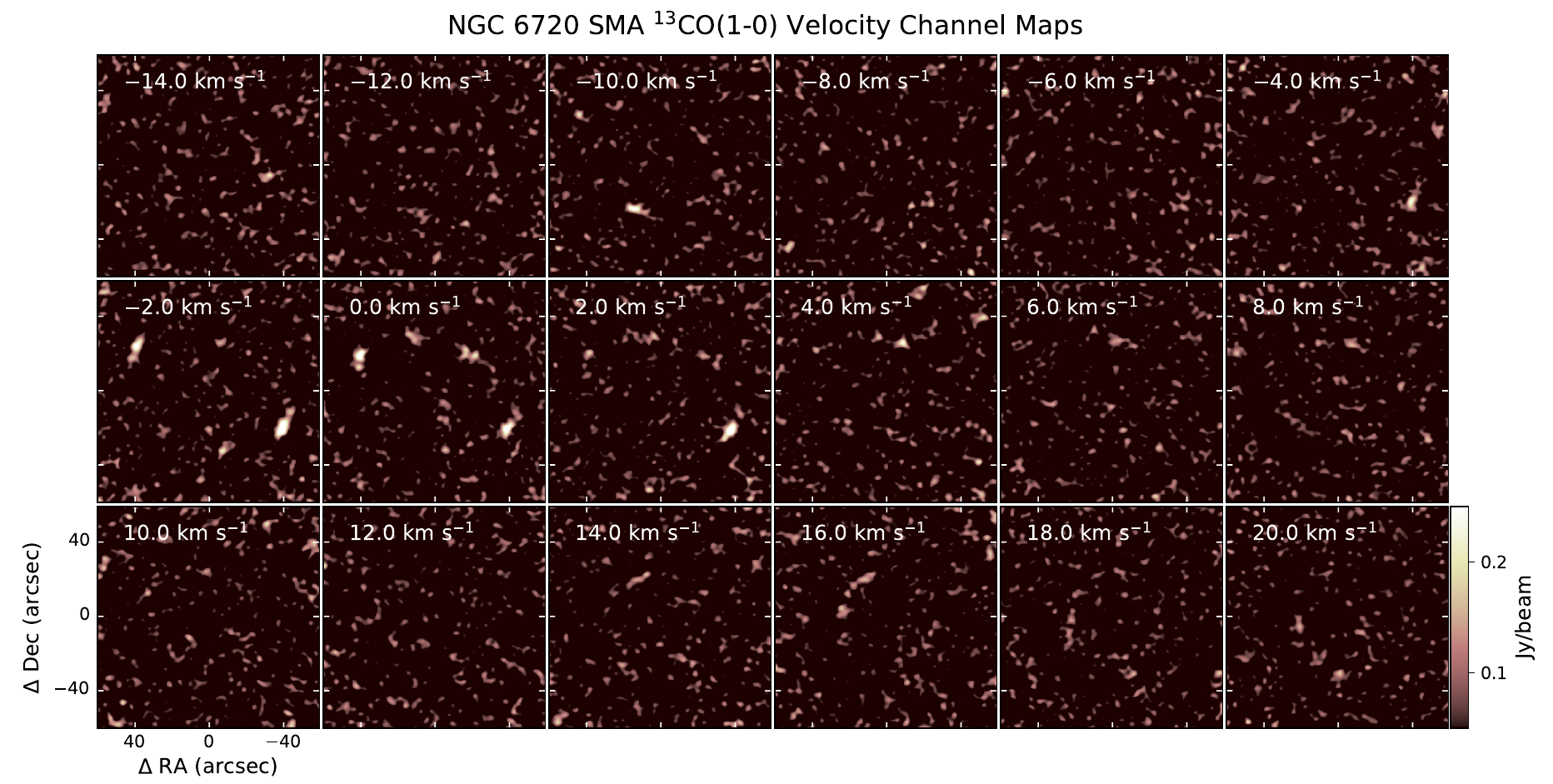}
\end{center}
\caption{SMA channel maps of $^{13}$CO(2--1) emission from NGC 3132. The synthesized beam size is 3.74$''\times$3.24$''$.}
\label{fig:13COchannelMaps}
\end{figure}

\begin{figure}[ht]
\begin{center}
\includegraphics[width=7.5in]{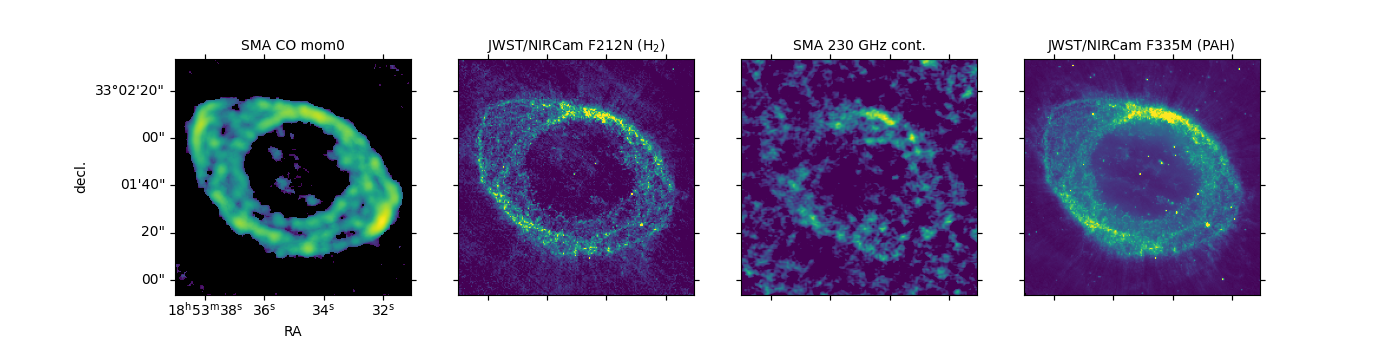}
\end{center}
\caption{Comparison of velocity-integrated (moment 0) image of $^{12}$CO(2--1) emission (left), JWST/NIRCam 2.12 $\mu$m H$_2$ image (left center)
SMA 230 GHz continuum image (right center), and JWST/NIRCam 3.35 $\mu$m PAH image (right).}
\label{fig:compare_mom0H2contPAH}
\end{figure}

The $^{12}$CO(2--1) channel maps (Fig.~\ref{fig:COchannelMaps}), P-V images (Fig.~\ref{fig:compare_cubes}), and $^{12}$CO(2--1) spectrum (Fig.~\ref{fig:COspectrum}) demonstrate that the bulk of the emission lies in the velocity range $-$26 km s$^{-1}$ to $+$16 km s$^{-1}$, with the brightest emission (including the aforementioned bright clumps along the major axis) found around the nebular systemic velocity of $\sim$$-$5 km s$^{-1}$. These regions generate the primary, central peak at $-$5 km s$^{-1}$ in the spatially integrated $^{12}$CO spectrum (Fig.~\ref{fig:COspectrum}). The ``shoulders'' (secondary peaks) in the spectrum, at $-$20 km s$^{-1}$ and $+$16 km s$^{-1}$, correspond to the two, nearly circular ends of the main  $^{12}$CO(2--1) emission velocity range (see channel maps at these velocities in Fig.~\ref{fig:COchannelMaps}).  

Indeed, overall --- ignoring the clumpy nature of the emission --- the  $^{12}$CO(2--1) ``surface'' revealed by the channel maps and P-V images (Figs.~\ref{fig:COchannelMaps},~\ref{fig:compare_cubes}) takes the form of a truncated ellipsoid that is viewed nearly along its symmetry axis. This 3D  $^{12}$CO(2--1) emission surface hence closely follows that of the optical shell, as inferred by \citet{ODell2007} on the basis of their multi-angle long-slit spectroscopy. Furthermore, the cold ($\sim$30-100 K) molecular gas traced by the $^{12}$CO(2--1) imaging is evidently confined to a thin, clumpy layer surrounding the ionized nebula. The individual $^{12}$CO(2--1) channel maps, particularly those near the systemic velocity ($-5$ km s$^{-1}$), demonstrate that the emission from this clumpy molecular gas layer has a thickness similar to the SMA beam, i.e., $\sim$4$''$, suggesting that the  $^{12}$CO(2--1) emitting layer is only marginally resolved. 


\subsection{Comparison with JWST infrared imaging}

Fig.~\ref{fig:compare_mom0H2contPAH} and Fig.~\ref{fig:compare_JWST_SMA} display comparisons, respectively, of velocity-integrated (moment 0) SMA  $^{12}$CO(2--1) and continuum images with the JWST/NIRCam images of 2.12 $\mu$m H$_2$ and 3.35 $\mu$m PAH emission and the JWST/MIRI 18 $\mu$m image of [S {\sc iii}] and ionized continuum emission.
Overall, the velocity-integrated  $^{12}$CO(2--1) emission morphology is strikingly similar to the H$_2$ and PAH emission morphologies (Fig.~\ref{fig:compare_mom0H2contPAH}). In particular, all three images display bright knots along the minor axis of the Ring, and the central, near-circular cavity imaged by JWST also appears as a  $^{12}$CO(2--1) emission cavity in the SMA imaging. However, in contrast to the minor-axis  $^{12}$CO(2--1) features --- which correspond closely to the brightest regions of near-IR emission from (hot) H$_2$ and small dust grains (PAHs) as imaged by JWST --- the bright clumps at the tips of the nebula's major axis in the  $^{12}$CO(2--1) moment 0 image have no clear counterparts in the near-IR H$_2$ or PAH imaging. These CO knots likely trace pockets of cold molecular gas that lie at the extremes of the Ring's ellipsoidal molecular envelope. The spatial relationship between the thin layer of cold molecular gas traced by the SMA imaging and the ionized gas imaged by JWST is further demonstrated by the contours of $^{12}$CO(2--1) emission at the nebular systemic velocity that are overlaid on the image of [S {\sc iii}] and ionized continuum emission in Fig.~\ref{fig:compare_JWST_SMA}. Evidently, the thin layer of cold molecular gas imaged in $^{12}$CO(2--1) by the SMA envelopes the ionized gas.

In the following subsections we describe additional insight gained into the central star's position and mass-loss history that can be gleaned from the comparison of the SMA mm-wave  $^{12}$CO(2--1) and JWST infrared images.

\subsubsection{Position of central star}

In their analysis of JWST imaging, \citet{Wesson2024} found that the central star of NGC~6720 appears to be offset by $\sim$2$"$ ($\sim$1500 au) to the SE of the central, nearly circular, nebular cavity that is apparent in the JWST imaging.  However, the narrow- to medium-band JWST infrared imaging only yields integrated emission morphologies as projected on the plane of the sky. Thanks to the combination of high spatial and spectral resolution of the molecular line imaging, we can revisit the question of the central star's (3D) position within the nebula. 

In Fig.~\ref{fig:CSPNposition}, we display the central star's position, as determined from JWST/NIRCam PAH image (Fig.~\ref{fig:CSPNposition}, panel {\it a}), overlaid on $^{12}$CO(2--1) emission-line maps extracted from the SMA data cube over narrow velocity ranges centered at the nebular systemic velocity and at the blueshifted and redshifted ``caps'' of the nearly pole-on truncated ellipsoid. This Figure demonstrates that the central star is, in fact, found at the geometric center of the elliptical  $^{12}$CO(2--1) shell as imaged at the systemic velocity (Fig.~\ref{fig:CSPNposition}, panel {\it c}). 
Furthermore, the star is offset by essentially identical $\sim$2$"$ angular displacements --- but in opposite directions --- with respect to the roughly circular blueshifted and redshifted  $^{12}$CO(2--1) emission regions that are symmetically displaced in velocity (Fig.~\ref{fig:CSPNposition}, panels {\it b} and {\it c}, respectively). These oppositely directed, symmetric angular displacements of the central star in the blueshifted and redshifted $^{12}$CO(2--1) images are purely the result of the $\sim$12$^\circ$ inclination of the ellipsoidal nebula's truncation symmetry axis with respect to our line of sight (see \S~\ref{sec:model}). 

We conclude that there is no significant (measurable) offset of NGC~6720's central, post-AGB star from the geometric center of the nebula, as defined in 3D by the $^{12}$CO(2--1) emission geometry in position-velocity space. The apparent off-center location of the central star obtained from JWST imaging  \citep{Wesson2024} can be attributed entirely to the star's projected displacement from the center of the nearly circular redshifted end of the ellipsoidal emission surface, combined with the relative deficit of obscuring, forward-facing molecular gas and dust to the northwest of the central star (note the lack of blueshifted  $^{12}$CO(2--1) to the NW of the central star in channel map velocities from $-$26 to $-$22 km s$^{-1}$, in Fig.~\ref{fig:COchannelMaps}). 

\subsubsection{High-velocity CO features}\label{sec:highVCO}

Fig.~\ref{fig:HVfeatures} displays the SMA moment 0  $^{12}$CO(2--1) image 
overlaid with blue and red contours representing
blueshifted and redshifted  $^{12}$CO(2--1) spanning velocity ranges $-54$ to $-36$
km s$^{-1}$ and $+34$ to $+52$ km s$^{-1}$, respectively, alongside the
JWST/NIRCam PAH image, with the same contours overlaid. This
Figure reveals that some (although perhaps not all) of the faint  $^{12}$CO(2--1) emission features seen projected within the central cavity are in fact 
fast-moving, molecule-rich polar knots or bullets with radial velocities of $\sim$45--50 km s$^{-1}$ relative to systemic.
Fig.~\ref{fig:HVfeatures} furthermore demonstrates that these compact, high-velocity molecular emission features have direct counterparts in 
the JWST images in the form of faint, filamentary structures seen projected in the Ring's interior, oriented along the major axis of the outer elliptically-shaped ring through the Ring's central cavity. 
The high-velocity  $^{12}$CO(2--1) emission features also correspond to the ``Lobe'' structures (also termed ``inner region features'') in the schematic model of the Ring proposed by \citet[][their Fig.~11]{ODell2013}. As that structural model suggested, the strong implication is that the high-velocity blue- and redshifted molecular emission arises in structures projected well in front of and behind the main Ring. Additional implications of these spatial correspondences are considered in \S~\ref{sec:OriginStory}.

\subsection{Total molecular mass and $^{12}$C/$^{13}$C isotopic ratio}

To estimate the total mass of molecular gas in the envelope of NGC~6720 from the SMA $^{12}$CO(2--1) data, we follow the methods described in \citet{Kastner2024}. In that treatment, estimates of the CO column densities $N_{\rm CO}$ are obtained from RADEX modeling \citep{VanderTak2007} assuming that the gas kinetic temperature $T_k$ is 30 K, the H$_2$ number density is $10^6$ cm$^{-3}$, and the $^{12}$CO(2--1) emission is optically thin. We adopt those same assumptions here, which yield a representative estimate of $N_{\rm CO} = 7\times10^{14}$ cm$^{-2}$ for a measured brightness temperature of 1.0 K \citep{Kastner2024}.  We note that the assumption $T_k = 30$ K is consistent with the $^{12}$CO excitation temperature range within NGC~6720 inferred from single-dish mapping \citep[i.e., 16--38 K; ][]{Bachiller1989}, and the assumption that the emission is optically thin is supported by the RADEX modeling. 

The mean surface brightness of  $^{12}$CO(2--1) in the moment 0 image of NGC~6720 is measured as 1.25 Jy km s$^{-1}$ beam$^{-1}$, or 4.7 K km s$^{-1}$, within a $\sim$$100'' \times 80''$ elliptical region encompassing the main ring. For the preceding assumptions for emitting region physical conditions, this corresponds to a mean CO column density of $N_{\rm CO} \sim 4\times10^{15}$ cm$^{-2}$. Adopting main ring dimensions of $1.1\times10^{18}$ cm $\times$ $7.6\times10^{17}$ cm, we then find that the total number of CO molecules is  $\sim$$3.3\times10^{51}$.  For the same range in CO:H$_2$ number ratio that \citet{Kastner2024} assumed for purposes of estimating the total molecular mass of NGC 3132 --- i.e., upper and lower limits of $10^{-4}$ and $10^{-5}$, respectively --- we obtain a total H$_2$ mass of between $\sim$0.05 $M_\odot$ and $\sim$0.5 $M_\odot$. 

Interestingly, this estimate for the plausible range of the total molecular mass of NGC~6720 is roughly three times larger than that inferred for NGC 3132 using the very same methods \citep{Kastner2024}. Our estimated lower limit on the total molecular mass of NGC~6720 is furthermore consistent with, albeit somewhat smaller than, the lower limit of 0.1 $M_\odot$ deduced by \citet{Bachiller1989} from their single-dish  $^{12}$CO(2--1) mapping data for an assumed CO:H$_2$ ratio of $10^{-4}$. The factor $\sim$2 discrepancy is likely due to those investigators' use of a spatially variable $T_k$, constrained by measurements of the variation of the $^{12}$CO(2--1)/$^{12}$CO(1--0) emission-line ratio across the nebula.

Although the SMA $^{13}$CO(2--1) data have low signal-to-noise ratio (\S \ref{sec:results}), we can nonetheless use these observations to obtain an estimate of the $^{12}$CO(2--1)/$^{13}$CO(2--1) line ratio and, hence, the $^{12}$C/$^{13}$C isotopic ratio in the NGC~6720 molecular gas. Single-dish observations of NGC~6720 at (RA, decl.) offsets from the central star of ($-$40$''$, $-$20$''$), a location of locally strong mm-wave CO emission, have yielded measurements of the CO isotopologue $J=2\rightarrow 1$ line ratio that range from $^{12}$CO(2--1)/$^{13}$CO(2--1) $\sim$ 9 \citep[][29$''$ beam]{Balser2002} to $\sim$22 \citep[][12$''$ beam]{Bachiller1997,Palla2000}. Integrating the SMA data cube at this same position within a circular aperture of diameter 12$''$, to approximate the beamsize used by \citet{Palla2000}, we obtain total intensity measurements of 154 Jy km s$^{-1}$ and 14 Jy km s$^{-1}$ for the $^{12}$CO(2--1) and $^{13}$CO(2--1) lines, respectively. Our result for the $^{12}$CO(2--1) line intensity is in excellent agreement with that obtained by \citet{Palla2000} at the same location; however, our $^{13}$CO(2--1) line intensity measurement is roughly twice that reported by those investigators, leading to a $^{12}$CO(2--1)/$^{13}$CO(2--1) ratio of $\sim$ 11, i.e., more consistent with the \citet{Balser2002} result. 

Since the frequencies of the $J=2\rightarrow 1$ lines only differ by a few percent, our result $^{12}$CO(2--1)/$^{13}$CO(2--1) $\sim$11 directly implies a $^{12}$C/$^{13}$C isotopic ratio of $\sim$11 (with an estimated uncertainty, due to calibration uncertainties, of $\sim$15\%), again assuming optically thin $^{12}$CO(2--1) emission. This result for NGC 6720's $^{12}$C/$^{13}$C isotopic ratio is consistent with estimates of $^{12}$C/$^{13}$C $\sim$12 and $\sim$11, as obtained from isotopologues of HCN and CN, respectively, by \citet{Ziurys2020}. As noted by those authors, such a relatively low $^{12}$C/$^{13}$C ratio might suggest that the AGB progenitor of NGC 6720 underwent hot bottom burning, but this is not expected for progenitor masses as low as that typically inferred for the Ring's progenitor \citep[$\sim$2 $M_\odot$;][and references therein]{Wesson2024}.

\section{A triaxial ellipsoid model}\label{sec:model}

\begin{figure}[ht]
\begin{center}
\includegraphics[width=5.5in]{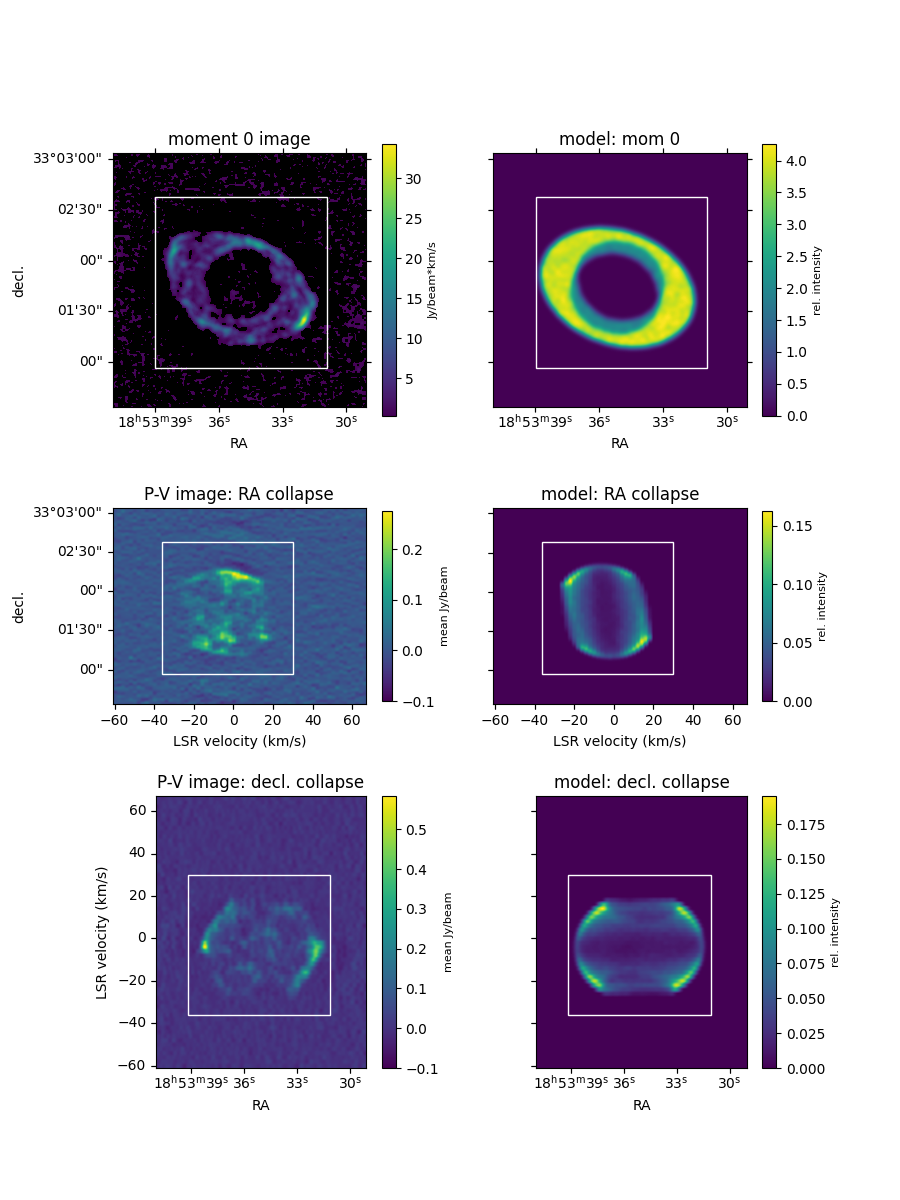}
\end{center}
\caption{{\it Left panels:}
Moment 0 and position-velocity images of  $^{12}$CO(2--1) emission obtained by collapsing the SMA data cube along each dimension. {\it Right panels:} The corresponding collapsed-cube views of the triaxial
  ellipsoid model described in \S \ref{sec:model} (right panels). The white square in each panel has dimensions of $80\times$80 kau$^2$, adopting the Gaia parallax distance to NGC~6720 (782 pc) and the CO envelope's extent along the $Z$ axis, where the latter is inferred from the truncated ellipsoid model described in \S \ref{sec:model}.}
\label{fig:compare_cubes}
\end{figure}

\begin{figure}[ht]
\begin{center}
\includegraphics[width=5in]{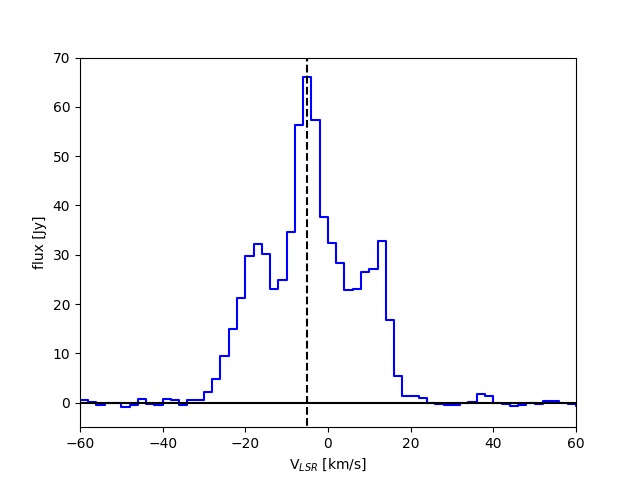}
\end{center}
\caption{Spatially integrated SMA $^{12}$CO(2--1) spectrum of NGC~6720.} 
\label{fig:COspectrum}
\end{figure}

\begin{figure}[]
\begin{center}
\includegraphics[width=5in]{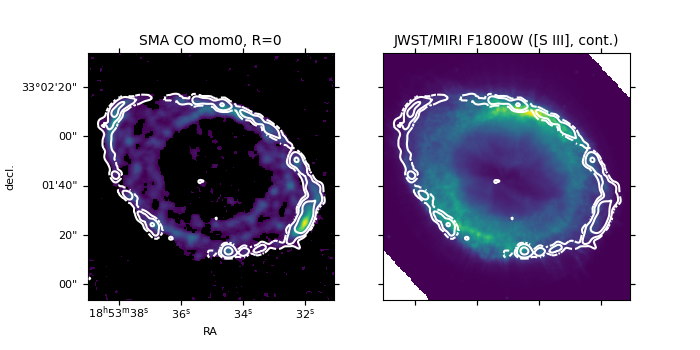}
\end{center}
\caption{Comparison of SMA moment 0  $^{12}$CO(2--1) image (left) with JWST/MIRI
F1800W image, covering $\sim$18 $\mu$m [S {\sc iii}] and continuum emission (right). Contours of the SMA $^{12}$CO(2--1) image isolating a 4 km s$^{-1}$ velocity interval centered on the nebular systemic velocity are overlaid on each panel, to demonstrate how the cold molecular gas traced by the SMA imaging envelopes the ionized gas imaged by JWST.}
\label{fig:compare_JWST_SMA}
\end{figure}

\begin{figure}[ht]
\begin{center}
\includegraphics[width=6.5in]{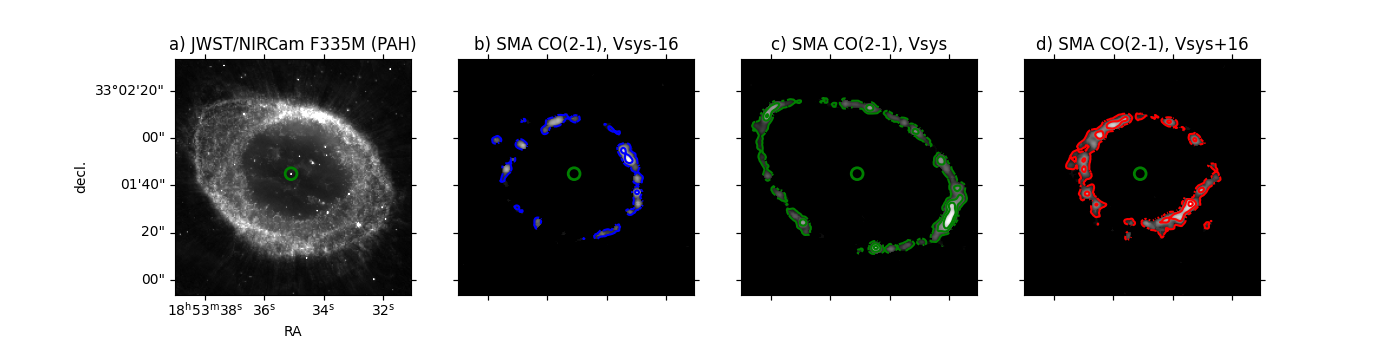}
\end{center}
\caption{Left panel (a):  JWST/NIRCam PAH image of NGC~6720, with the position of the central star (FK5 coordinates 18:53:35.102, +33:01:45.00) highlighted by the green circle. Next three panels, (b)--(d): SMA $^{12}$CO(2--1) images, in greyscale, integrated over velocity intervals that are (b) blueshifted by $\sim$16 km s$^{-1}$ with respect to the nebular systemic velocity of $-6$ km s$^{-1}$ (i.e., LSR velocities of $-$24 to $-20$ km s$^{-1}$); (c) centered on the nebular systemic velocity (i.e., LSR velocities of $-$8 to $-4$ km s$^{-1}$); and (d) redshifted by $\sim$16 km s$^{-1}$ with respect to systemic (i.e., LSR velocities of $+$8 to $+12$ km s$^{-1}$). The images are overlaid with blue, green, and red contours at levels of 0.5, 2.5 and 5.0 Jy km s$^{-1}$ in panels (b), (c), and (d), respectively; the position of the central star is indicated by the green circle in each panel.}
\label{fig:CSPNposition}
\end{figure}

\begin{figure}[ht]
\begin{center}
\includegraphics[width=6.5in]{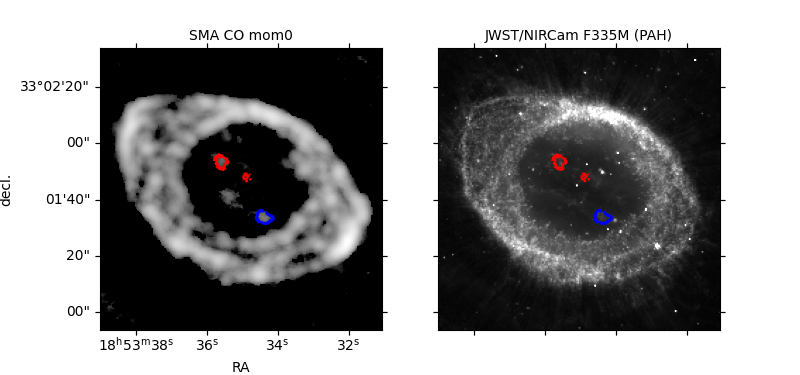}
\end{center}
\caption{Left: SMA moment 0   $^{12}$CO(2--1) image 
  overlaid with blue and red contours representing
  blueshifted and redshifted CO spanning velocity ranges $-54$ to $-36$
  km s$^{-1}$ and $+34$ to $+52$ km s$^{-1}$, respectively. Right:
  JWST/NIRCam PAH image, with the same contours overlaid.}
\label{fig:HVfeatures}
\end{figure}

As noted (\S~\ref{sec:COmorph}), NGC~6720 clearly displays a truncated ellipsoidal emission morphology in the SMA $^{12}$CO(2--1) ``collapsed-cube'' P-V images (see Fig.~\ref{fig:compare_cubes}, middle left and bottom left panels), and \citet{ODell2007} deduced a triaxial ellipsoidal geometry for the ionized core region of the nebula from the P--V behavior of their long-slit optical spectroscopy data. Hence, to reproduce the essential features of the $^{12}$CO(2--1) emission in position-velocity space, we have formulated a geometric model consisting of an expanding triaxial ellipsoid model that is oriented and truncated (nearly) along the line of sight. 
The basic modeling methodology follows that of \citet{Kastner2024}. Briefly, the geometrical shape --- here, a truncated ellipsoid --- is formulated in $XYZ$ space, with the $XY$ plane corresponding to the plane of the sky and $Z$ to velocity. The dimensions and orientation of the shape are constrained by the observed angular extent ($XY$) and expansion velocity ($Z$) of the nebula, as revealed by the SMA $^{12}$CO(2--1) data. The intensities across the surface of the ellipsoid are set to random (arbitrary) values between 0 and 1, to crudely mimic a clumpy medium. The resulting surface shape is then convolved with Gaussians approximating the interferometer beam and spectrometer line spread function. Finally, the convolved model surface is transformed to the SMA data cube P-V space. 

Through trial and error, we find that a model ellipsoid with $X:Y:Z$ axial ratios of 1.0 : 0.71: 0.82 that is viewed within $\sim$$2^\circ$ of the $Z$ axis, is rotated around $Z$ through a position angle of $\sim$70$^\circ$ (measured E from N), and is truncated by a highly elongated, more highly inclined ellipsoid (see below), provides a reasonable match to the data. In this model, the expansion velocity is set to a maximum of 30 km s$^{-1}$ along the longest ellipsoid axis  and the model dynamical age is set to 6000 yr (for a distance of 782 pc). The implicit assumption is that the entire CO envelope was ejected with an ellipsoidal velocity field in a relatively short-lived event, $\sim$6000 yr ago. As we have not performed a rigorous fit of models to the data, the dimensions of the basic ellipsoid surface --- and, hence, derived quantities such as expansion velocity and dynamical age --- are uncertain at the $\sim$10\% level. We also note that the onset of nebular photoionization by the central post-AGB star likely has accelerated the expansion of the molecular envelope, such that the dynamical age may be an underestimate of the elapsed time since the onset of envelope ejection. Nonetheless, we see that for this adopted dynamical age and expansion velocity, the ellipsoid has $X \times Y$ dimensions of $1.1\times10^{18}$ cm $\times$ $7.6\times10^{17}$ cm, as observed (see Fig.~\ref{fig:compare_cubes}, top panels). 

\begin{figure}[ht]
\begin{center}
\includegraphics[width=3in]{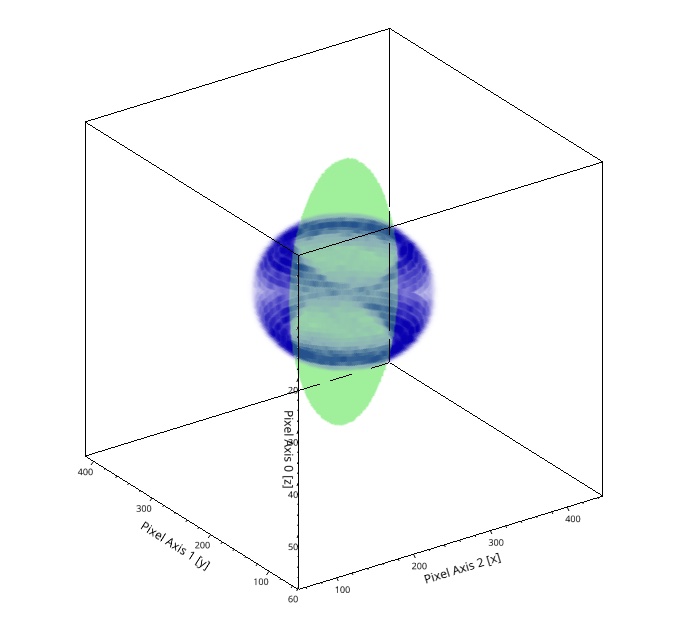}
\includegraphics[width=3in]{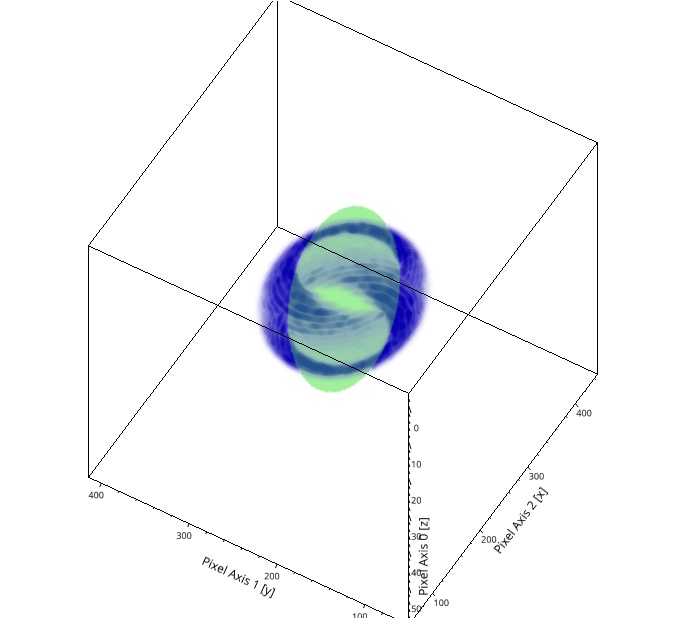}
\includegraphics[width=3in]{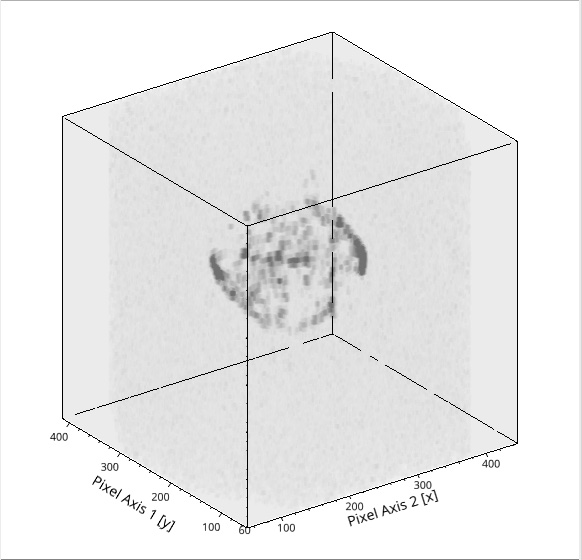}
\includegraphics[width=3in]{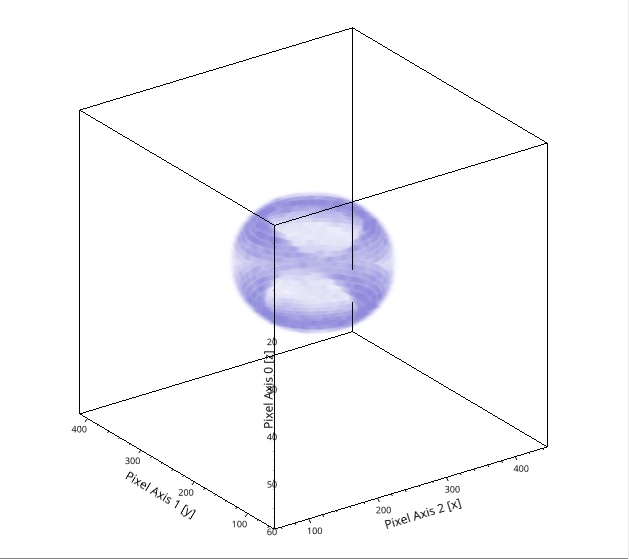}
\includegraphics[width=3in]{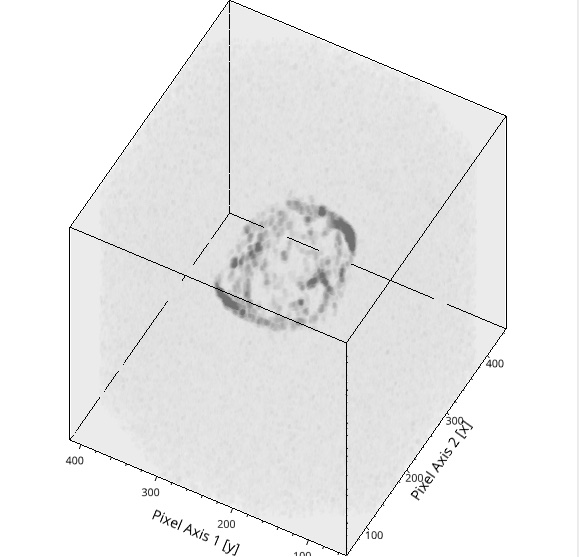}
\includegraphics[width=3in]{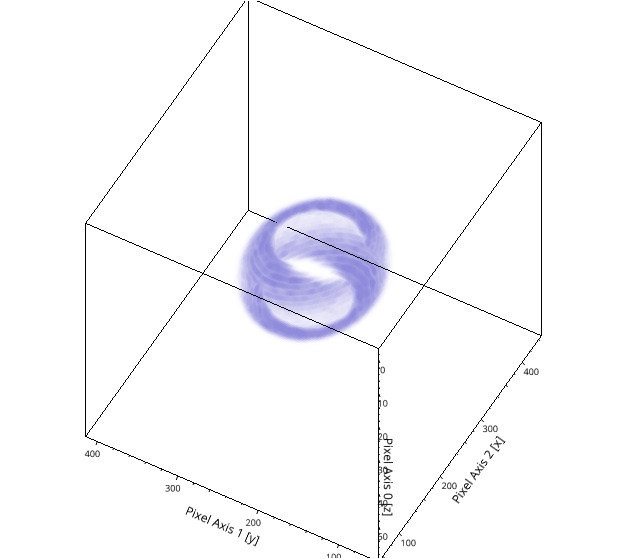}
\end{center}
\caption{{\it Top panels:} Two views of the truncated ellipsoidal model components (see \S \ref{sec:model}). The truncated triaxial ellipsoid surface is shown in blue, and the intersecting inclined, axisymmetric ``mask'' ellipsoid is shown in green. {\it Middle and lower panels:} Comparison of example oblique views of the SMA  $^{12}$CO(2--1) data cube (left
  panels) and the same views of the triaxial ellipsoid model data cube (right panels). The dimensions of the cube are $\sim$150$''$ along the R.A., decl. ($x, y$) axes and $-$60 to +60 km s$^{-1}$ along the velocity (z) axis, and the voxels are scaled such that the three dimensions have the same physical length (see \S \ref{sec:model}).} 
\label{fig:obliqueViews}
\end{figure}

Truncation (nearly) along the $Z$ direction is then introduced to account for the fact that the molecular envelope is ``open-ended'' along $Z$, i.e., that the emission abruptly terminates at radial velocities of $\sim$$\pm$20 km s$^{-1}$ with respect to systemic (i.e., at LSR velocities $-24$ km s$^{-1}$ and $+14$ km s$^{-1}$; Fig.~\ref{fig:COchannelMaps}). We find these truncations occur at distance of $\sim$$8\times10^{17}$ cm along the $Z$ direction. Furthermore, the  $^{12}$CO(2--1) emission appears nearly circular in channel maps at and near these velocity extremes, and the centers of the near-circular ``caps'' are offset by $\sim$4$''$ (e.g., Fig.~\ref{fig:CSPNposition}). To model the structure of these nearly circular, spatially offset ellipsoid truncations --- and to explore the potential relationship of these structures to the high-velocity features seen projected within the main ring of the nebula (\S \ref{sec:highVCO}) --- we introduce an axially symmetric ellipsoidal ``mask;'' the surfaces of intersection of this elongated, tilted ellipsoid with the main triaxial ellipsoid are set to zero intensity. While we caution that the geometry of this second, ellipsoidal mask is not tightly constrained, we find that a mask with semiminor axis (radius) $2.3\times10^{17}$ cm and major axis (height) $1.8\times10^{18}$ cm and is inclined by 12$^\circ$ at PA 160$^\circ$ (measured E from N) with respect to the line of sight renders a truncated ellipsoid whose dimensions and truncations closely match those observed in the SMA $^{12}$CO(2--1) data. Oblique views of the truncated ellipsoid and ellipsoidal ``mask'' are presented in  Fig.~\ref{fig:obliqueViews} (top panels). 

\begin{figure}[ht]
\begin{center}
\includegraphics[width=6.5in]{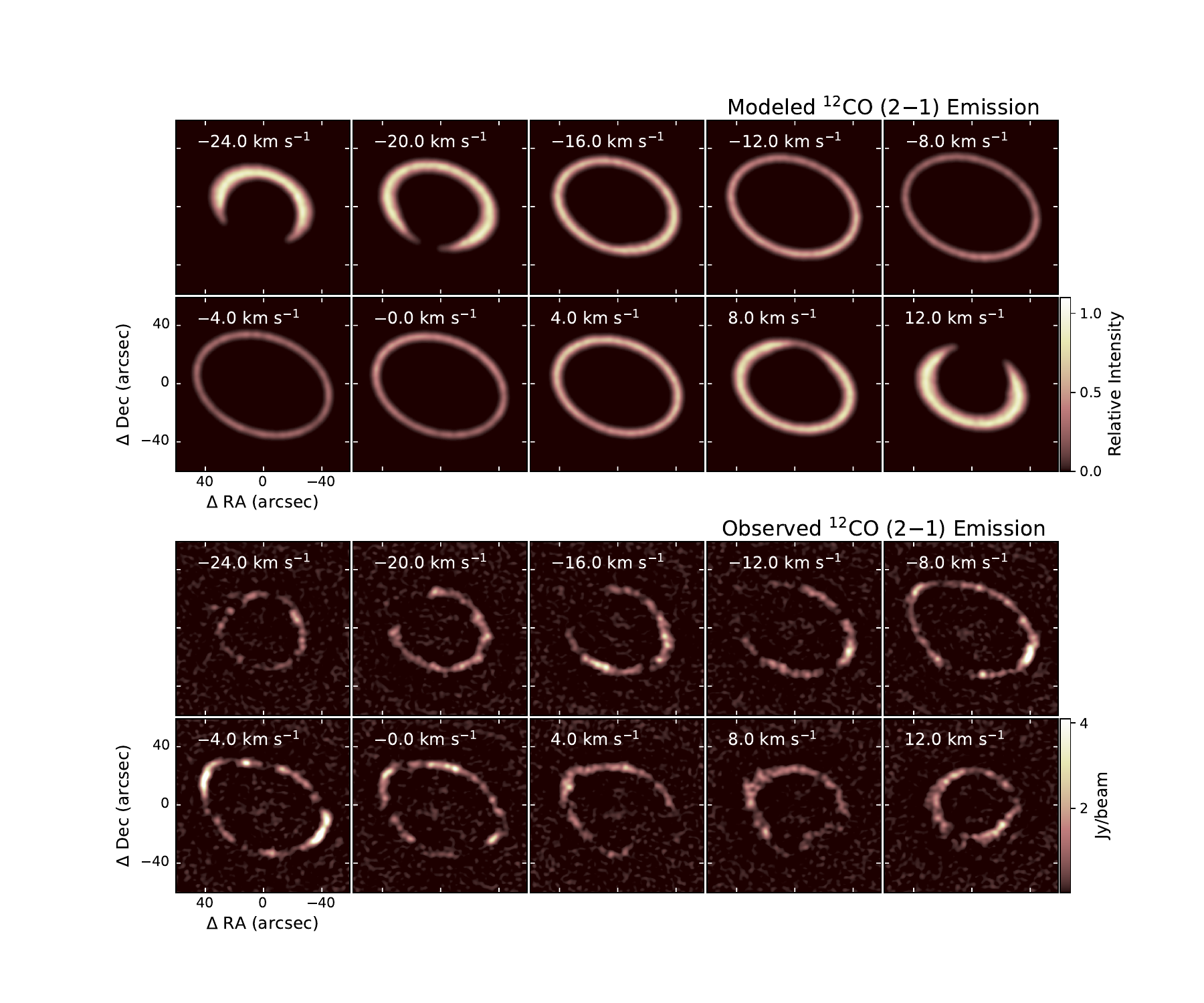}
\end{center}
\caption{Comparison of channel maps for the triaxial ellipsoid model (top panels) with the equivalent SMA channel maps of $^{12}$CO(2--1) emission from NGC~6720 (bottom panels). The images are 4 km s$^{-1}$ in width and span the velocity range over which emission from the main eilipsoidal shell is detected (see Fig.~\ref{fig:COchannelMaps}).}
\label{fig:COchannelMapComp}
\end{figure}

Comparisons of data and model are presented in the form of moment 0 and P-V images in Fig.~\ref{fig:compare_cubes}, 
representative oblique views in Fig.~\ref{fig:obliqueViews}, and velocity channel maps in Fig.~\ref{fig:COchannelMapComp}. Figs.~\ref{fig:compare_cubes} and~\ref{fig:obliqueViews} demonstrate that the ellipsoid model closely traces the extent, overall shape and morphology, and inclination of the CO envelope, and can reproduce the basic morphologies of the  $^{12}$CO(2--1) emission as revealed in the channel map sequence (Fig.~\ref{fig:COchannelMapComp}). 
However, the geometric model described here is only formulated to reproduce the approximate dimensions of the (truncated ellipsoidal) molecular envelope, with no attempt to simulate its small-scale structures. Hence, the model does not capture the myriad clumps and knots observed in the SMA maps of  $^{12}$CO(2--1) emission from the Ring. In particular, the Ring displays prominent, bright clumps along its major axis at approximately systemic velocity. These features --- evident in Fig.~\ref{fig:compare_cubes}
--- appear to be part of a point-symmetric system of bright knots (see middle left and lower left panels of Fig.~\ref{fig:compare_cubes}).

An important aspect of the model's predictive power is the ability to use the dynamical age of NGC~6720 (6000 yr) and expansion velocity along the $Z$ direction (24 km s$^{-1}$), as inferred from the model, to visualize this (velocity) dimension of the data cube in terms of spatial scale, as is of course straightforward for $X$ and $Y$ directions (given the well-determined distance to the nebula). 
The resulting 3D spatial scaling is illustrated in the form of white squares of dimension 80 kau $\times$ 80 kau superimposed on the moment 0 and P-V images in Fig.~\ref{fig:compare_cubes}. The same scaling is used in the oblique views presented in  Fig.~\ref{fig:obliqueViews}. It is evident from the extent of $^{12}$CO(2--1) emission relative to the fiducial white squares in Fig.~\ref{fig:compare_cubes}, and from inspection of the data cube views in Fig.~\ref{fig:obliqueViews}, that the line-of-sight extent of the nebula's molecular envelope is similar to that of its major axis (and larger than its minor axis) as seen projected on the sky.

\newpage

\section{Discussion}

\subsection{The 3D structure, inclination, and dynamical age of the Ring Nebula}

Modeling of the SMA $^{12}$CO(2--1) data (\S~\ref{sec:model}) demonstrates that the intrinsic structure of NGC~6720's main, bright ring is that of a prolate, triaxial ellipsoid viewed nearly (within $\sim$2$^\circ$ of) pole-on, with the ellipsoid symmetrically truncated along the polar (line-of-sight) direction. We have modeled these truncations as tracing the intersection of the main, prolate ellipsoid and a second, more highly elongated and slightly ($\sim$12$^\circ$) inclined ellipsoid. The small but significant deviation from a pole-on orientation of the ellipsoid's polar truncations, combined with a deficit of emission from the forward-facing (blueshifted) regions of the truncated ellipsoid, explains the apparent offset of the PN's central star from the (projected) geometric center of the central nebular cavity reported by \citet[][]{Wesson2024}. 

In our model, the $\sim$12$^\circ$ inclination of the surface of ellipsoid truncation, rather than the (nearly pole-on) ellipsoid itself, is mainly responsible for generating the ``skew'' observed in the $^{12}$CO(2--1)  P--V images (Fig.~\ref{fig:compare_cubes}). This modest inclination is consistent with that inferred for the nebular core region by \citet[][$i \sim 6^\circ$ and $i \sim 13^\circ$, respectively]{ODell2007,ODell2013}, but is much smaller than the polar axis inclination of 30$^\circ$ that was estimated by \citet{Bryce1994} and \citet{Guerrero1997} and, most recently, was adopted for the twin-cone model described in \citet[][]{Wesson2024}. However, we note that the \citet{Bryce1994} and \citet{Guerrero1997} inclination estimates were based on the assumption that the Ring's inner ($\sim$150$''$ diameter) halo and bright, central ring (respectively) are intrinsically circular --- whereas the SMA data definitively demonstrate that the main ring (at least) departs significantly from circular symmetry. \citet{Guerrero1997} also cited the appearance of tilts in P--V images obtained along specific spectrometer slit directions as supporting a larger inclination, but these tilt measurements may be biased somewhat by the Ring's complex, point-symmetric system of emission knots.

The modeling indicates that the expansion velocity of the nebula is $\sim$30 km s$^{-1}$ and $\sim$20 km s$^{-1}$ along the Ring's (projected) major and minor axes, respectively, and that the dynamical age of the nebula's molecular envelope is $\sim$6000 yr. This dynamical age estimate is consistent, given the uncertainties, with the kinematic age estimate obtained from comprehensive long-slit spectroscopy of the Ring's ionized emission lines \citep[$\sim$5200 yr;][]{ODell2007}; both estimates are somewhat larger than the dynamical age of $\sim$4000 yr obtained from feature proper motions via multiepoch HST imaging by \citet[][]{ODell2013}. Our new results for the Ring's dynamical age and plane-of-sky expansion velocities can be tested via additional multi-epoch HST and/or JWST imaging.

\subsection{The stratified structure of the Ring: molecular gas enveloping ionized gas}

The comparison of velocity-resolved SMA mapping of  $^{12}$CO(2--1) with JWST imaging furthermore conclusively demonstrates that the cold molecular gas traced by CO resides in a thin shell that lies exterior to (envelopes), rather than partially embedded within, the ionized gas. 
Previously, when considering the origin of the Ring Nebula's molecular gas reservoir, it had been assumed that a significant mass of molecular gas is present within the main, ionized elliptical ring of NGC~6720 \citep[e.g.,][and references therein]{vanHoof2010}. However, this assumption was based on the nebula's integrated-intensity  near-IR H$_2$ and mm-wave CO emission morphologies, which only yield the projection of the molecular emission on the plane of the sky. With the benefit of the $\sim$2 km s$^{-2}$ resolution of the SMA  $^{12}$CO(2--1) mapping, we now see that the molecular gas whose emission appears projected within the nebular interior in fact sits just in front of and just behind the ionized gas. The stratified structure of the Ring that emerges from the comparison of SMA and JWST data --- wherein the ionized gas surrounded by a thin envelope of molecular gas and dust --- serves as a strong argument for a ``primordial'' AGB star origin for the molecular gas in the Ring and Ring-like PNe, rather than an {\it in situ}, post-AGB origin, such as formation of H$_2$ in recombining ionized gas and/or on the surfaces of dust grains embedded in the ionized gas \citep[as proposed by][]{vanHoof2010}.
  
\subsection{NGC 6720 and NGC 3132: two Rings, one origin story}\label{sec:OriginStory}

The SMA data reveal that, just as in the case of NGC 3132, some of the faint, filamentary
features seen projected in the Ring's interior in JWST imaging are in fact
fast-moving, molecular knots or bullets located near the nebular polar axis. The molecular knots in NGC~6720 display higher radial velocities ($\sim$45--50 km s$^{-1}$ relative to systemic) than those observed in NGC 3132 ($\sim$25 km s$^{-1}$ relative to systemic). The NGC~6720 high-velocity features hence appear analogous to the high-velocity molecular ``bullets'' observed in the young planetary nebula BD $+$30$^\circ$ 3639 \citep{Bachiller2000}, which are located well outside the main BD $+$30$^\circ$ 3639 nebula. As noted in \S~\ref{sec:highVCO}, the strong implication is that the blueshifted and redshifted high-velocity gas (and dust) structures projected within the Ring are also, in fact, located (respectively) well in front of and behind NGC~6720's central cavity, consistent with the basic model described by \citet[][]{ODell2013}. These high-velocity knots perhaps trace the tips of the surface represented by the elongated ellipsoid ``mask'' that serves to truncate of the Ring's main, prolate ellipsoid molecular envelope, in the model described in \S~\ref{sec:model} and visualized in the top two panels of Fig.~\ref{fig:obliqueViews}.

More generally, the SMA interferometric  $^{12}$CO(2--1) mapping studies of the Ring (NGC~6720) and the Southern Ring (NGC~3132) reported here and in \citet{Kastner2024}, respectively, serve to underline the remarkable parallels between these two molecule-rich PNe. They are evidently not only morphologically but also structurally very similar. Both PNe display  intrinsic molecular envelope geometries best characterized as clumpy ellipsoids, with high-velocity features that trace polar ejecta seen projected within their main rings.  Given their respective dynamical ages ($\sim$3700 yr for NGC 3132 vs.\ $\sim$6000 yr for NGC~6720) --- and notwithstanding the fact that the molecular envelope of NGC~6720 appears to be a factor $\sim$3 more massive than that of NGC 3132 --- the two PNe appear to represent an evolutionary (growth) sequence that directly traces the rapid formation of a molecule-rich PN back to its origins in ellipsoidal, axisymmetric AGB ejecta. Indeed, as speculated by \citet{Bachiller1993}, such equatorially-enhanced ellipsoidal density structures may be the rule, rather than the exception, for Ring-like PNe.
  
Moreover, the combination of an ellipsoidal envelope with truncated, circular ends and high-velocity molecular bullets along that same (polar) axis appears to point to the influence of an interacting companion star on the geometry and rate of mass loss from NGC~6720's AGB star progenitor. As in the case of NGC 3132 \citep{Kastner2024}, it appears likely that the relatively sudden mass ejection event that generated NGC~6720's ellipsoidal molecular envelope likely terminated the progenitor star's asymptotic giant branch evolution, and the envelope was subsequently disrupted by fast, collimated polar outflows resulting from interaction between the progenitor and its companion(s).

\section{Summary and Conclusions}

The Ring (planetary) Nebula, NGC~6720, is undoubtedly among the most
photographed objects in the night sky. However, its intrinsic
structure, as well as its formation and evolution, have been fiercely
debated for well over a century. Here, we have presented the first-ever
interferometric imaging of molecular line emission from NGC~6720, in
the form of Submillimeter Array (SMA) observations of CO
$J=2\rightarrow 1$ emission. The SMA  $^{12}$CO(2--1) data, obtained at
$\sim$3$''$ spatial resolution and 2 km s$^{-1}$ velocity resolution,
provide an unprecedentedly detailed, 3D view of the Ring's massive,
clumpy molecular envelope. Our main conclusions are as follows:

\begin{itemize}

\item The morphology of the velocity-integrated SMA  $^{12}$CO(2--1) image
closely resembles those of near-IR H$_2$ and PAH emission as revealed
in recent JWST/NIRCam imaging of NGC~6720. However, the SMA  $^{12}$CO(2--1) imaging also
betrays the presence of pockets of cold molecular gas within the
Ring's exterior regions that are invisible to JWST's IR cameras. 
These molecule-rich knots are distributed in 
point-symmetric fashion with respect to the nebula's polar axis. 

\item The intrinsic 3D envelope geometry, as revealed
by the SMA CO data, is that of a truncated, triaxial ellipsoid viewed
nearly pole-on. Specifically, we find that a model truncated ellipsoid with axial ratios of 1.0 : 0.71 : 0.82, viewed within $\sim$$2^\circ$ of pole-on and with its surface of truncation inclined by $\sim$12$^\circ$, provides a reasonable match to the data. The modeling indicates that the molecular envelope dynamical age is $\sim$6000 yr, with an expansion velocity of $\sim$30~km~s$^{-1}$ along the longest ellipsoid axis (which is oriented at position angle $\sim$70$^\circ$, measured E from N). 

\item This basic 3D structure as well as the dynamical age of the PN inferred from the modeling are consistent with the basic model postulated for NGC~6720 by \citet{ODell2007,ODell2013} on the basis of optical spectroscopy and HST imaging. However, the combination of nearly pole-on ellipsoidal geometry and high-velocity polar knots projected within the main ring revealed by the SMA observations is very difficult (if not impossible) to reconcile with the ``tipped twin-cone'' model that was invoked by \citet{Wesson2024} to interpret the Ring's morphology as observed in JWST imaging. 

\item The SMA  $^{12}$CO(2--1) data furthermore demonstrate that the molecular gas is
confined to a system of CO-bright knots or clumps forming a
geometrically thin layer that immediately surrounds the ionized gas
imaged by HST and JWST, with the hot post-AGB remnant of the nebular progenitor star
positioned at the precise geometric center of this ellipsoidal
shell. 

\item The lack of detectable  $^{12}$CO(2--1) within the Ring's ionized interior region in the
SMA data serves as strong evidence that its clumpy molecular gas
reservoir originated in the envelope of the progenitor asymptotic
giant branch (AGB) star, as opposed to forming {\it in situ} within
recombining ionized nebular gas. Indeed, the SMA  $^{12}$CO(2--1) data confirm that
some of the faint, filamentary features seen projected near the
central star in the Ring's interior in HST and JWST imaging are in
fact fast-moving ($\sim$45--50 km s$^{-1}$), polar knots or bullets 
located well in front of and behind the central star. 

\item The Ring's molecular envelope likely represents the ``fossil'' remnant of a
relatively short-lived mass loss event $\sim$6000 yr ago that terminated the
progenitor star's AGB evolution. It appears this ellipsoidal envelope of
AGB ejecta was subsequently disrupted by fast, collimated polar
outflows or jets resulting from interactions between the progenitor
and one or more companion stars. 

\item The two-stage, AGB-terminating, binary-driven mass ejection process just described for the Ring Nebula constitutes the same basic scenario that 
we \citep{Kastner2024} invoked to explain the CO emission structure and kinematics of the Southern Ring Nebula (NGC 3132), as revealed by SMA mapping, 
with the main differences being that NGC 3132 is dynamically younger (age $\sim$3700 yr) and somewhat less massive, and its density distribution appears more equatorially concentrated than that of NGC 6720.
Interferometric CO observations of additional molecule-rich, Ring-like planetary nebulae could establish whether the formation of most if not all molecule-rich, Ring-like planetary nebulae proceeds in similar fashion: i.e., ellipsoidal ejection of the molecule-rich AGB envelope, followed by faster, collimated, bipolar flows resulting in envelope punctures or truncations and systems of molecular bullets and/or filaments along the polar directions. 

\end{itemize}

\vspace{.5in}

{\it Acknowledgements.} Research by J.K., D.R., and P.M.B. on the
molecular content of planetary nebulae is supported by NSF grant AST--2206033 to RIT. R.S.'s contribution to the research described
here was carried out at the Jet Propulsion Laboratory, California
Institute of Technology, under a contract with NASA, and
funded in part by NASA via ADAP awards, and multiple HST
GO awards from the Space Telescope Science Institute.
The Submillimeter Array is a joint project between the Smithsonian Astrophysical Observatory and the Academia Sinica Institute of 
Astronomy and Astrophysics and is funded by the Smithsonian Institution and the Academia Sinica. We recognize that Maunakea is a 
culturally important site for the indigenous Hawaiian people; we are privileged to study the cosmos from its summit.

\facility{SMA}

\end{document}